\documentclass[twocolumn,prb,amsmath,amssymb,showpacs,superscriptaddress,floatfix]{revtex4}

\usepackage{graphicx}

 \DeclareMathOperator{\Real}{\mathrm{Re}}
 \DeclareMathOperator{\Imag}{\mathrm{Im}}
 \DeclareMathOperator{\sgn}{sgn}
 \DeclareMathOperator{\Tr}{Tr}

\begin{document}

 \title{Josephson effect due to the long-range odd-frequency triplet superconductivity\\
        in SFS junctions with N\'eel domain walls}

 \author{Ya.~V.~Fominov}
 \email{fominov@landau.ac.ru}
 \affiliation{L.~D.~Landau Institute for Theoretical Physics RAS, 119334 Moscow, Russia}

 \author{A.~F.~Volkov}
 \email{volkov@tp3.rub.de}
 \affiliation{Theoretische Physik III, Ruhr-Universit\"{a}t Bochum, D-44780 Bochum, Germany}
 \affiliation{Institute of Radioengineering and Electronics of the Russian Academy of Sciences, 103907 Moscow, Russia}

 \author{K.~B.~Efetov}
 \affiliation{Theoretische Physik III, Ruhr-Universit\"{a}t Bochum, D-44780 Bochum, Germany}
 \affiliation{L.~D.~Landau Institute for Theoretical Physics RAS, 119334 Moscow, Russia}

\date{22 March 2007}

\begin{abstract}
We consider a SFS Josephson junction made of two superconductors S and a multidomain ferromagnet F with an in-plane
magnetization. We assume that the neighboring domains of the ferromagnet are separated by N\'eel domain walls. An
odd-frequency triplet long-range component of superconducting correlations arises in the domain walls and spreads into
the domains over a long distance of the order $\xi_T = \sqrt{D/2\pi T}$, where $D$ is the diffusion coefficient (dirty
limit is implied). We calculate the contribution of this component to the Josephson current in the situation when
conventional short-range components exponentially decay over the thickness of the F layer and can be neglected. In the
limit when the thickness of the F layer is much smaller than the penetration length of the long-range component, we find
that the junction is in the $\pi$ state. We also analyze a correction to the density of states due to the long-range
triplet component.
\end{abstract}


% 74.45.+c Proximity effects; Andreev effect; SN and SNS junctions

% 74.78.-w Superconducting films and low-dimensional structures
% 74.78.Fk Multilayers, superlattices, heterostructures

% 74.50.+r Tunneling phenomena; point contacts, weak links, Josephson effects (for SQUIDs, see 85.25.Dq; for Josephson
% devices, see 85.25.Cp; for Josephson junction arrays, see 74.81.Fa)

% 75.70.-i Magnetic properties of thin films, surfaces, and interfaces (for magnetic properties of nanostructures, see
% 75.75.+a)
% 75.70.Kw Domain structure (including magnetic bubbles)

\maketitle

\section{Introduction}

The past decade was marked by a rapid growth of interest to the study of hybrid superconductor-ferromagnet (SF)
structures (see, for example, reviews \onlinecite{Buzdin,LP,BVE_review}). New physical phenomena arising in these
systems originate from a nontrivial interplay of competing orders in superconductors and ferromagnets. On one side,
electron-electron interactions lead in superconductors to formation of Cooper pairs consisting of two electrons with
opposite spins. On the other side, the exchange interaction in ferromagnets tends to align the electron spins parallel
to each other. In SF structures these two types of interactions are spatially separated and can coexist despite much
greater value of the exchange energy $h$ in comparison with the superconducting gap $\Delta$.

Due to the proximity effect \cite{deGennes} the superconducting correlations penetrate into the ferromagnet in SF
structures. The condensate wave function $f$ penetrates into the ferromagnet with a uniform magnetization $\mathbf M$
over a distance of the order of the ``exchange length'' $\xi_h =\sqrt{D/h}$.\cite{Buzdin,LP} The condensate wave
function decays in F in a nonmonotonic way as $f(x) \sim \exp (-x/\xi_h)\cos (x/\xi_h)$: it oscillates in space and
decreases exponentially. This nonmonotonic behavior of $f(x)$ leads, in particular, to a $\pi$ state in SFS Josephson
junctions\cite{Buzdin,LP,Golubov,Ryazanov,Kontos,Blum,Strunk,Sellier} characterized by a negative critical current $I_c$
in the Josephson current-phase relation $I(\varphi) = I_c \sin\varphi$.

If the magnetization in the ferromagnet is nonuniform a new phenomenon becomes possible: a triplet component of the
condensate wave function $f$ (generally speaking, the condensate wave function is a matrix in the particle-hole and spin
space) arises in the SF system.\cite{BVE_review} This triplet component is an odd function of the Matsubara frequency
$\omega$ (while the conventional BCS singlet component of $f$ is an even function of $\omega$) and spreads in the
ferromagnet over a long distance of the order of $\xi_T =\sqrt{D /2\pi T}$. The existence of this long-range triplet
odd-frequency component in SF structures with an inhomogeneous magnetization was predicted in Ref.~\onlinecite{BVElong}
and further discussed in Ref.~\onlinecite{Kadigrobov}. Unlike the triplet component in superfluid $^3$He and in
Sr$_2$RuO$_4$, this odd-frequency triplet component corresponds to \textit{s}-wave pairing and, hence, is symmetric in
the momentum space. Therefore it is not destroyed by scattering on nonmagnetic impurities and survives in the dirty
limit. We call this component the long-range triplet component (LRTC). Historically, the odd-frequency triplet pairing
was conjectured in 1974 by Berezinskii\cite{Berez} as a possible mechanism for superfluidity in $^3$He but this
conjecture was not confirmed experimentally.

There is a significant amount of experimental data that may be interpreted as manifestation of the LRTC in SF
systems.\cite{Petrashov1,Giordano,Giroud,Petrashov2,Pena,Petrashov3,Greek,Keizer,Petrashov06} Of particular importance
are the experiments on SFS systems in which a long-range phase coherence of the condensate wave functions was observed
in ferromagnets with a length considerably exceeding the penetration length of the singlet component
$\xi_h$.\cite{Pena,Keizer,Petrashov06} In principle, this long-range phase coherence can be due to the LRTC but, for
unambiguous identification of the LRTC, further experimental and theoretical studies are very important. In particular,
the theory of the LRTC was developed for specific types of the inhomogeneous magnetization in ferromagnets, which do not
exhaust all possible types of the magnetic structures in real samples.

In Refs.~\onlinecite{BVElong}, \onlinecite{Kadigrobov}, and \onlinecite{Volkov06} the LRTC was studied in SF systems
with a Bloch-type magnetic structure [the magnetization vector $\mathbf M(x)$ lies in the $y$-$z$ plane parallel to the
SF interface and rotates with increasing $x$]. The amplitude and the penetration length of the LRTC induced in the
ferromagnet have been calculated in Refs.~\onlinecite{BVElong} and \onlinecite{Volkov06}. Under certain conditions the
LRTC penetrates the ferromagnet over a long distance of order $\xi_T$. As shown in Ref.~\onlinecite{Volkov06}, where a
SFS structure with a conical ferromagnet was studied, the LRTC may decay in a nonmonotonic way provided the cone angle
exceeds a certain value. In Refs.~\onlinecite{VBE} and \onlinecite{BVEmanif} a multilayered SF structure was
investigated in which the magnetization vector $\mathbf M(x)$ has fixed but different orientations in different
ferromagnetic layers. This structure is also similar to a Bloch-type domain structure.

At the same time, it is well known that the domain structure in a ferromagnet can be not only of the Bloch type but also
of the N\'eel type.\cite{Aharoni} A SF system with a N\'eel-type magnetic spiral structure [the $\mathbf M(y)$ vector
lies in the $y$-$z$ plane parallel to the SF interface and rotates with increasing $y$] was studied theoretically in
Ref.~\onlinecite{Eschrig2}. This magnetic structure may be regarded as an infinite N\'eel wall. It was shown that in
this case the LRTC does not arise in the system. However, this statement is valid only for the case of a uniformly
rotating $\mathbf M(y)$ vector. In a more realistic situation of a ferromagnet with magnetic domains separated by N\'eel
walls, the LRTC arises at the domain walls and decays inside the domains over a long distance.\cite{VFE} Another source
of the triplet component in SF structures was considered in Ref.~\onlinecite{Eschrig}, where the SF interface was
assumed to be spin-active. The Josephson effect in a S/HM/S junction (HM is a ferromagnetic half-metal) was studied in
this paper and it was shown that the critical Josephson current has a maximum at low, but nonzero, temperature.

In the present paper we consider a SFS Josephson junction with the N\'eel domain structure in the ferromagnetic layer
and calculate the Josephson critical current $I_c$ in this system. Up to now there have been no theoretical
investigations of this problem. In Refs.~\onlinecite{BVEcrcur}, \onlinecite{VBE}, \onlinecite{BVEmanif}, and
\onlinecite{Volkov06} the critical current $I_c$ was calculated for a magnetic structure similar to the Bloch type. A
SFS Josephson junction with a rotating $\mathbf M(x)$ and with the thickness of the F film $d$ of order $\xi_h$ was
studied in Ref.~\onlinecite{BVEcrcur}. It was shown that the rotation of the magnetization vector $\mathbf M(x)$ leads
to the appearance of the LRTC and to a suppression of the $\pi$ state in the Josephson junction. The situation in a
multilayered SF structure with noncollinear magnetizations in the F layers is more complicated. The sign of $I_c$
depends on the chirality, that is, on whether the $\mathbf M$ vector rotates in the same direction in the whole system
or it oscillates with respect to the $z$ axis.\cite{VBE,BVEmanif} In a SFS junction with a conical ferromagnet of
thickness $d$ much greater than $\xi_h$, the critical current $I_c$ is due to the LRTC and the sign of $I_c$ depends on
the cone angle.\cite{Volkov06}

Note that in ferromagnets with a multidomain structure the LRTC does not arise if the thickness of the domain walls is
very small. In this case the magnetization vectors $\mathbf M$ are collinear and only singlet and short-range triplet
components are induced due to the proximity effect. The critical current in a SFS junction with such a structure was
calculated in Ref.~\onlinecite{Blanter}. The domain structure also results in a suppression of the $\pi$ state due to an
effective averaging of the exchange field. Moreover, if the exchange field $\mathbf h$ changes its sign over a scale
shorter than $\max( \xi_h, l )$ (where $l$ is the mean free path), long-range effects arise even in the absence of the
LRTC (the exchange field is effectively averaged out). This case is realized provided an antiferromagnet (AF) is used in
SFS junctions instead of a ferromagnet. The Josephson current in S/AF/S junctions was studied theoretically in
Refs.~\onlinecite{Gorkov} and \onlinecite{Barash} and experimentally (see Ref.~\onlinecite{Ovsyann} and references
therein).

\begin{figure}
 \includegraphics[width=7cm]{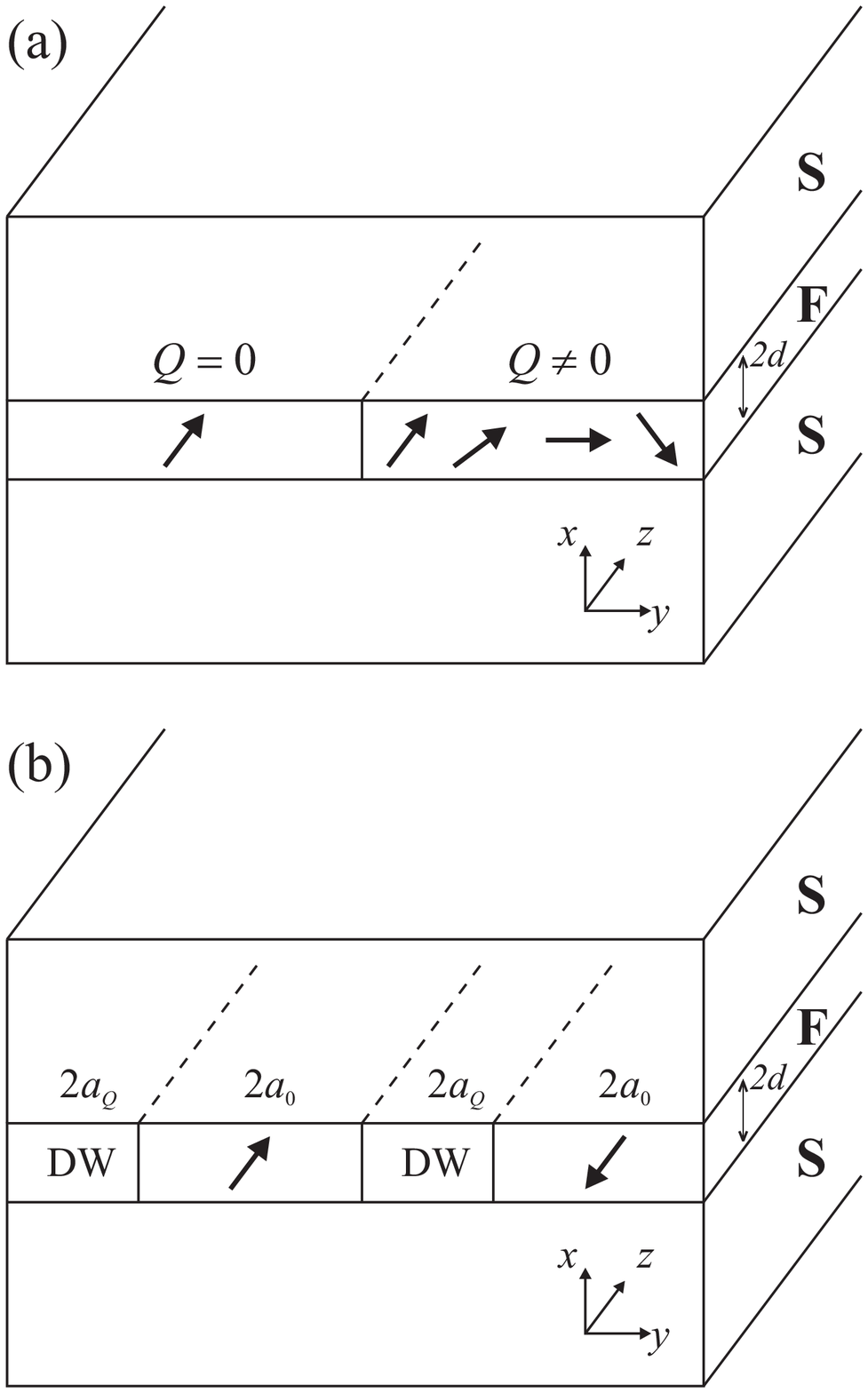}
\caption{SFS junctions considered in the paper. (a)~The domain ($y<0$) and the region with rotating magnetization
($y>0$) in the F layer are half-infinite. (b)~Multidomain F layer. Depending on the relative orientation of rotating
magnetizations in the neighboring domain walls, we distinguish the cases of positive and negative chirality ($Q$ has the
same or opposite sign in the neighboring domain walls, respectively).}
 \label{fig:system}
\end{figure}

We consider a domain structure in a thin F film, where domains with antiparallel in-plane magnetizations are separated
by the N\'eel walls (while the magnetization does not change across the thin F film). This domain structure is realized
in real ferromagnetic films.\cite{Aharoni} The $y$-$z$ plane is chosen to be parallel to the SF interfaces (see
Fig.~\ref{fig:system}). We show that the LRTC arises at the N\'eel domain walls and decays exponentially away from the
domain walls and the SF interfaces over a long distance $\xi_T$. We calculate the Josephson current due to the LRTC and
find that its sign corresponds to the $\pi$ junction. The mechanism of the $\pi$ junction in our case is related to
$\pi/2$ phase shifts at the SF interfaces, while the LRTC does not oscillate inside the F layer (in contrast to the
short-range component). We also study modifications of the density of states in the F film due to the LRTC.

The paper is organized as follows. In Sec.~\ref{sec:basic}, we formulate the Usadel equations and the corresponding
boundary conditions, investigate the main features of the long-range triplet superconducting component that appears due
to the presence of N\'eel domain walls, and find the Josephson current due to this component. The analysis is made for
the simplified model with only one half-infinite region with rotating magnetization. In Sec.~\ref{sec:multi}, we
consider the realistic case of the multidomain F layer using the results of Sec.~\ref{sec:basic}. In Sec.~\ref{sec:dos},
we study the correction to the density of states due to the triplet component. This section is an addition to our
previous paper Ref.~\onlinecite{VFE}. Finally, we discuss limitations of our model in Sec.~\ref{sec:discuss} and present
our conclusions in Sec.~\ref{sec:concl}.

\section{Basic equations. Josephson effect in the case of a half-infinite domain} \label{sec:basic}

We consider a ferromagnet ($0<x<2d$) sandwiched between two bulk superconductors. We assume that the in-plane exchange
field $\mathbf h(y)$ in the F layer proportional to the magnetization $\mathbf M$, depends on $y$ as follows: $\mathbf h
(y) = h (0,0,1)$ at $y<0$ and $\mathbf h (y)=h (0,\sin \alpha (y),\cos \alpha (y))$ with $\alpha (y)=Qy $ at $y>0$. This
means that the magnetization vector $\mathbf M$ is oriented along the $z$ axis at $y<0$ and rotates in the $y$-$z$ plane
at $y>0$. The region with the rotating magnetization models the N\'eel domain wall. This structure is shown in
Fig.~\ref{fig:system}(a) and contains only one half-infinite domain and one half-infinite region with rotating
magnetization. Then we shall use the obtained results to describe a realistic multidomain structure depicted in
Fig.~\ref{fig:system}(b).

In order to calculate the Josephson current, we need to find the condensate Green functions in the ferromagnet induced
due to the proximity effect. We consider the dirty limit, which means, in particular, that $h\tau \ll 1$, where $\tau$
is the momentum relaxation time due to elastic scattering.

In the dirty limit, the system is described by the Usadel equation for the matrix Green function $\check g$ which is a
$4\times 4$ matrix in the Gor'kov-Nambu and spin spaces. The Usadel equation for the case of an inhomogeneous
magnetization was written in Refs.~\onlinecite{BEL} and \onlinecite{BVE_review}. However, we redefine the Green function
used in Refs.~\onlinecite{BEL} and \onlinecite{BVE_review} (and also in our previous paper Ref.~\onlinecite{VFE})
introducing a new matrix function $\check g_\mathrm{new} = \check V \check g \check V^\dagger$ with the transformation
matrix
\begin{equation}
\check V = \exp\left( i\frac\pi 4 (\hat\tau_3-\hat\tau_0) \hat\sigma_3 \right),
\end{equation}
where $\hat\tau_i$ and $\hat\sigma_i$ are the Pauli matrices in the Gor'kov-Nambu and spin spaces, respectively. This
transformation was proposed in Ref.~\onlinecite{IF}, and below we shall use the new Green function $\check
g_\mathrm{new}$ omitting the subscript for brevity. The convenience of the new definition is that the Usadel equation
for the new Green function possesses the explicit symmetry with respect to rotations of the exchange field $\mathbf h$:
\begin{gather}
D\nabla \left( \check g \nabla \check g \right) - \omega \left[ \hat\tau_3 \hat\sigma_0, \check g \right]- i \left[
\hat\tau_3 (
\mathbf h \hat{\boldsymbol\sigma}), \check g \right] - \left[ \check\Delta , \check g \right] =0, \label{usadel} \\
\check\Delta = \left( \hat\tau_1 \Real\Delta - \hat\tau_2 \Imag\Delta \right) \hat\sigma_0, \notag
\end{gather}
where $\omega$ is the Matsubara frequency, $\mathbf h = h \mathbf n$, $\mathbf n = (0, \sin\alpha(y), \cos\alpha(y))$
and $\alpha (y)=0$ at $y< 0$ while $\alpha (y)=Qy$ at $y> 0$. We assume that the diffusion coefficients $D$ for
electrons with spins up and down are equal to each other (this is correct provided the exchange energy $h$ is much less
than the Fermi energy $\varepsilon_F$).

The Usadel equation (\ref{usadel}) is written in the general form taking into account both the superconductivity and
magnetism. In the SFS junction, the pair potential $\Delta$ is nonzero only in the S layers, while the exchange field
$\mathbf h$ is nonzero only in the F layer. There is no attractive interaction between electrons in the F layer, hence
$\Delta=0$. However, the condensate (Gor'kov) functions are finite in the F region due to the boundary conditions at the
SF interface.

In the general case, the Green function has the following components:
\begin{equation} \label{g_components}
\check g = \hat\tau_3 \left( g_0 \hat\sigma_0 + \mathbf g \hat{\boldsymbol\sigma} \right) + \hat\tau_1 \left( f_0
\hat\sigma_0 + \mathbf f \hat{\boldsymbol\sigma} \right) + \hat\tau_2 \left( \bar f_0 \hat\sigma_0 + \bar{\mathbf f}
\hat{\boldsymbol\sigma} \right).
\end{equation}
In the bulk of a normal metal, only the $g_0$ component is present. The superconducting correlations are described by
the scalar anomalous components $f_0$ and $\bar f_0$, while the vector components ($\mathbf g$, $\mathbf f$,
$\bar{\mathbf f}$) are due to the ferromagnetism.

The superconducting correlations described by the $f$ components (nondiagonal in the Nambu-Gor'kov space) are assumed to
be weak due to a finite interface transparency. In the considered case of weak proximity effect ($|f| \ll 1$), the Green
function acquires the form
\begin{equation}
\check g = \hat\tau_3 \hat\sigma_0 \sgn\omega + \check F,
\end{equation}
where the anomalous part can be written as
\begin{equation}
\check F = \hat\tau_1 \hat f + \hat\tau_2 \hat{\bar f}
\end{equation}
with matrices in the spin space
\begin{gather}
\hat f = f_0 \hat\sigma_0 + \mathbf f \hat{\boldsymbol\sigma}, \\
\hat{\bar f} = \bar f_0 \hat\sigma_0 + \bar{\mathbf f} \hat{\boldsymbol\sigma}.
\end{gather}
Equation (\ref{usadel}) can be linearized and brought to the form
\begin{equation} \label{usadelM}
\nabla^2 \check F - 2 k_\omega^2 \check F - i k_h^2 \sgn\omega \left\{ \hat\tau_0 (\mathbf n \hat{\boldsymbol\sigma}),
\check F \right\} =0,
\end{equation}
where $\omega = \pi T (2n+1)$, $k_\omega^2 = |\omega |/D$, $k_h^2 = h /D$, and the braces denote the anticommutator.

The anomalous part of the Green function in the bulk of the superconductor with the superconducting phase $\varphi$ is
$\check F_S(\varphi) = (\hat\tau_1 \cos\varphi - \hat\tau_2 \sin\varphi) \hat\sigma_0 f_S$, where
\begin{equation}
f_S = \frac\Delta{\sqrt{\omega^2 +\Delta^2}}.
\end{equation}
We intend to find the Josephson current at the phase difference $\varphi$ between the two superconducting banks.
Assuming the phases of the left and right superconductors to be $-\varphi/2$ and $\varphi/2$, respectively, we write the
boundary conditions for $\check F$ at the SF interfaces:
\begin{equation} \label{BC}
\frac{\partial \check F}{\partial x} = \mp \frac{\check F_S(\mp\varphi/2)}{\gamma_b} ,
\end{equation}
where the two signs correspond to the left ($x=0$) and right ($x=2d$) SF interfaces, respectively. Here $\gamma_b = R_b
\sigma$, while $R_b$ is the interface resistance per unit area and $\sigma$ is the conductivity of the ferromagnet. This
boundary condition follows from the Kupriyanov-Lukichev ones\cite{Kupriyanov} provided two assumptions are satisfied:
(1)~the proximity effect is weak (i.e., $\gamma_b /\xi_h \gg 1$) and (2)~the bulk solution $\check F_S$ in the
superconductors is unperturbed and valid up to the interface (i.e., $\gamma_b /\xi_S \gg \sigma/\sigma_S$, where
$\xi_S=\sqrt{D_S/\Delta}$ and $\sigma_S$ are the coherence length and the conductivity in the S banks).

The technical problem with Eq.\ (\ref{usadelM}) is that this is a two-dimensional partial-derivative differential
equation. However, we can employ a trick similar to the one proposed in Ref.~\onlinecite{VFE}, which allows us to make
the Fourier transformation over $x$, reducing the problem to only one dimension $y$.

The equations for the $\hat f$ and $\hat{\bar f}$ functions split and for the $\hat f$ function we obtain
\begin{gather}
\nabla^2 \hat f - 2 k_\omega^2 \hat f - i k_h^2 \sgn\omega \left\{ (\mathbf n \hat{\boldsymbol\sigma}), \hat f
\right\} =0, \\
\left. \frac{\partial \hat f}{\partial x} \right|_{x=0,2d} = \mp \frac{f_S \cos\frac\varphi 2}{\gamma_b} \hat\sigma_0.
\end{gather}
The $\hat{\bar f}$ function obeys the same equation, although the boundary conditions are different:
\begin{gather}
\nabla^2 \hat{\bar f} - 2 k_\omega^2 \hat{\bar f} - i k_h^2 \sgn\omega \left\{ (\mathbf n \hat{\boldsymbol\sigma}),
\hat{\bar f} \right\} =0, \\
\left. \frac{\partial \hat{\bar f}}{\partial x} \right|_{x=0,2d} = -\frac{f_S \sin\frac\varphi 2}{\gamma_b}
\hat\sigma_0.
\end{gather}

The functions $\hat f$ and $\hat{\bar f}$ are defined for $0\leqslant x \leqslant 2d$. The function $\hat f$ is even
with respect to the center of the F layer, while $\hat{\bar f}$ is odd. We can continue the functions to the whole $x$
axis: for $\hat f$ we do it periodically and for $\hat{\bar f}$ --- antiperiodically, obtaining continuous functions in
both cases. Due to the boundary conditions both the functions have cusps at $x= 2dN$ with integer $N$. The boundary
conditions (\ref{BC}) producing the cusps can be incorporated into Eq.\ (\ref{usadelM}) itself as $\delta$-functional
terms (this situation is similar to the standard quantum-mechanical problem with $\delta$-functional potential producing
a cusp of the wave function):
\begin{multline} \label{usadelM1}
\nabla^2 \hat f - 2 k_\omega^2 \hat f - i k_h^2 \sgn\omega \left\{ (\mathbf n \hat{\boldsymbol\sigma}), \hat f \right\}
\\ = -\hat\sigma_0 \frac{2 f_S \cos\frac\varphi 2}{\gamma_b} \sum_{N=-\infty}^\infty \delta (x-2dN) ,
\end{multline}
\begin{multline} \label{usadelM1a}
\nabla^2 \hat{\bar f} - 2 k_\omega^2 \hat{\bar f} - i k_h^2 \sgn\omega \left\{ (\mathbf n \hat{\boldsymbol\sigma}),
\hat{\bar f} \right\}
\\ = -\hat\sigma_0 \frac{2 f_S \sin\frac\varphi 2}{\gamma_b} \sum_{N=-\infty}^\infty (-1)^N \delta (x-2dN) .
\end{multline}
Now, instead of solving Eq.\ (\ref{usadelM}) at $0\leqslant x \leqslant 2d$ with the boundary conditions (\ref{BC}), we
have to solve Eqs.\ (\ref{usadelM1}) and (\ref{usadelM1a}) at all $x$.

The Fourier transformation
\begin{equation}
f(k,y) =\int_{-d}^d dx \exp(-ikx) f(x,y)
\end{equation}
in Eqs.\ (\ref{usadelM1}) and (\ref{usadelM1a}) should be performed over ``bosonic'' wave vectors $k_n$ and
``fermionic'' wave vectors $\bar k_n$ for the periodic function $\hat f$ and antiperiodic function $\hat{\bar f}$,
respectively:
\begin{equation}
k_n = \frac\pi{2d}2n,\qquad \bar k_n =\frac\pi{2d} (2n+1).
\end{equation}
We obtain
\begin{gather}
\frac{\partial^2 \hat f}{\partial y^2} - (k_n^2 + 2 k_\omega^2) \hat f - i k_h^2 \sgn\omega \left\{ (\mathbf n
\hat{\boldsymbol\sigma}), \hat f \right\} = -\frac{2 f_S \cos\frac\varphi 2}{\gamma_b} \hat\sigma_0, \label{usadelM2} \\
\frac{\partial^2 \hat{\bar f}}{\partial y^2} - (\bar k_n^2 + 2 k_\omega^2) \hat{\bar f} - i k_h^2 \sgn\omega \left\{
(\mathbf n \hat{\boldsymbol\sigma}), \hat{\bar f} \right\} = -\frac{2 f_S \sin\frac\varphi 2}{\gamma_b} \hat\sigma_0.
\end{gather}

The two equations are similar and we may consider only one of them, say Eq.\ (\ref{usadelM2}) for the $\hat f$ function.
Then the result for the $\hat{\bar f}$ function can immediately be obtained by substituting $k_n \mapsto \bar k_n$ and
$\cos\frac\varphi 2 \mapsto \sin\frac\varphi 2$.

At $y>0$ the function $\alpha (y)$ is $y$-dependent, while at $y<0$ we have $\alpha =0$. In the region of positive $y$
one can exclude the $y$-dependence from Eq.\ (\ref{usadelM2}) with the help of a rotation
\begin{equation} \label{Utrans}
\hat f =\hat U \hat f_u \hat U^+,
\end{equation}
where $\hat U =\exp \bigl( i \hat\sigma_1 \alpha (y)/2 \bigr)$. As a result, we get ($y>0$)
\begin{multline} \label{usadelM3}
\frac{\partial^2 \hat f_u}{\partial y^2}- \left( k_n^2+\frac{Q^2}2 + 2 k_\omega^2 \right) \hat f_u +\frac{Q^2}2
\hat\sigma_1 \hat f_u \hat\sigma_1 +i Q \bigl[ \hat\sigma_1, \frac{\partial \hat f_u}{\partial y} \bigr] \\
-i k_h^2 \sgn\omega \bigl\{ \hat\sigma_3, \hat f_u \bigr\} = -\frac{2 f_S \cos\frac\varphi 2}{\gamma_b} \hat\sigma_0
\end{multline}
in terms of the new function $\hat f_u(k,y)$ (the square brackets denote the commutator). The same equation is valid for
$y<0$ if we put $Q=0$:
\begin{equation} \label{usadelM4}
\frac{\partial^2 \hat f_u}{\partial y^2} -\left( k_n^2 + 2 k_\omega^2 \right) \hat f_u -i k_h^2 \sgn\omega \bigl\{
\hat\sigma_3, \hat f_u \bigr\}  =-\frac{2 f_S \cos\frac\varphi 2}{\gamma_b} \hat\sigma_0.
\end{equation}

The original functions $\hat f$ and $\partial \hat f /\partial y$ are continuous at $y=0$. Therefore, the rotated
functions obey the following boundary conditions at $y=0$:
\begin{align}
\hat f_u (-0) &= \hat f_u (+0),  \label{MC1} \\
\frac{\partial \hat f_u (-0)}{\partial y} &= \frac{\partial \hat f_u (+0)}{\partial y} +i \frac Q2 \left[ \hat\sigma_1 ,
\hat f_u \right] . \label{MC2}
\end{align}

Thus, we have to solve the linear matrix Eqs.\ (\ref{usadelM3}) ($y>0$) and (\ref{usadelM4}) ($y<0$) of the second order
with the boundary conditions (\ref{MC1}) and (\ref{MC2}) at $y=0$. We can represent the solution in the form
\begin{equation} \label{f(y)}
\hat f_u =\hat{\mathcal F}(Q) \theta (y)+\hat{\mathcal F}(0)\theta (-y)+\delta \hat f_u,
\end{equation}
where $\theta$ is the Heaviside step function and the constants $\hat{\mathcal F}(Q)$ and $\hat{\mathcal F}(0)$ are the
homogeneous solutions of Eqs.\ (\ref{usadelM3}) and (\ref{usadelM4}) at $y=\pm\infty$. The matrices $\hat{\mathcal F}$
have the form
\begin{equation} \label{F}
\hat{\mathcal F} = \hat\sigma_0 \mathcal F_0 +\hat\sigma_3 \mathcal F_3,
\end{equation}
where
\begin{align}
\mathcal F_0 (Q) &= \frac{2 f_S \cos\frac\varphi 2 \left( k_n^2+Q^2+2 k_\omega^2 \right)}{\gamma_b \mathcal D(Q)},
\label{F_0} \\
\mathcal F_3 (Q) &= -\frac{4i f_S \cos\frac\varphi 2 k_h^2 \sgn\omega}{\gamma_b \mathcal D(Q)}, \label{F_3}
\end{align}
and
\begin{equation}
\mathcal D(Q)= \left( k_n^2+Q^2+2 k_\omega^2 \right) \left( k_n^2+2 k_\omega^2 \right)+4 k_h^4.
\end{equation}

The correction $\delta \hat f_u(k,y)$ obeys the same Eqs.\ (\ref{usadelM3}) and (\ref{usadelM4}) without the right-hand
side. It has the form
\begin{equation} \label{delta-f}
\delta \hat f_u = \hat\sigma_0 f_0 + \hat\sigma_3 f_3 + \hat\sigma_2 f_2.
\end{equation}
The first term in Eq.\ (\ref{delta-f}) is the singlet component. The second term is the triplet component with zero $z$
projection of the Cooper-pair spin. This component arises even in the case of a homogenous magnetization of the
ferromagnet and decays in the F film at the short distance $\xi_h$. The last term in Eq.\ (\ref{delta-f}) is the triplet
component with the spin projection $\pm 1$. It arises in the case of an inhomogeneous magnetization and decays over a
long distance of the order $\xi_T$. The functions $f_i(k,y)$ in Eq.\ (\ref{delta-f}) can be represented as a sum of
eigenfunctions of Eqs.\ (\ref{usadelM3}) and (\ref{usadelM4}), i.e.,
\begin{align}
f_i (y) &= \sum_l A_{il} \exp \bigl( -\kappa_l (Q) y \bigr)\quad \text{at } y>0  \label{A}, \\
f_i (y) &= \sum_l B_{il} \exp \bigl(  \kappa_l (0) y \bigr)\quad \text{at } y<0  \label{B}.
\end{align}
The inverse decay lengths $\kappa_l(Q)$ are the eigenvalues of Eqs.\ (\ref {usadelM3}) and (\ref{usadelM4}) (without the
right-hand side). The equation for $\kappa_l(Q)$ has the form ($l=1,2,3$)
\begin{multline}
\left[ \left( \kappa_l^2 -k_n^2 -Q^2 -2 k_\omega^2 \right)^2 + 4 \left( Q\kappa_l \right)^2 \right] \left( \kappa_l^2
-k_n^2 -2 k_\omega^2 \right) \\
+4 k_h^4 \left( \kappa_l^2 -k_n^2 -Q^2 -2 k_\omega^2 \right) = 0.
\end{multline}
We assume that the exchange length is the shortest length in the problem:
\begin{equation} \label{Assum}
k_h^2 \gg k_n^2, Q^2, k_\omega^2 .
\end{equation}
Then, the eigenvalues $\kappa_l$ consist of two ``short-range'' values
\begin{equation}
\kappa_\pm \approx \left( 1\mp i \sgn\omega \right) k_h
\end{equation}
and one ``long-range'' value
\begin{equation}
\kappa_L (Q) \approx \sqrt{k_n^2+Q^2+2 k_\omega^2}.
\end{equation}
At $y<0$ we have the same $\kappa_l$ with $Q=0$.

Calculating the corresponding eigenvectors and matching solutions (\ref{A}) and (\ref{B}) with the help of the boundary
conditions (\ref{MC1}) and (\ref{MC2}), we find the coefficients $A_{il}$ and $B_{il}$. This simple but cumbersome
calculation is similar to the one presented in Ref.~\onlinecite{VFE}. In the considered limit of a small exchange length
[see Eq.\ (\ref{Assum})], the coefficients $A_{2L}$ and $B_{2L}$ that describe the LRTC are the largest ones,
\begin{equation} \label{A1LB1L}
A_{2L} \approx B_{2L} \approx \frac{Q \mathcal F_3}{\kappa_Q +\kappa_0} ,
\end{equation}
where for brevity we have denoted
\begin{equation}
\kappa_Q \equiv \kappa_L(Q),\quad \kappa_0 \equiv \kappa_L(0).
\end{equation}
In the limit (\ref{Assum}) the function $\mathcal F_3 (Q)$ has a simple form
\begin{equation} \label{F0}
\mathcal F_3 \approx -\frac{i f_S \cos\frac\varphi 2 \sgn\omega}{\gamma_b k_h^2} .
\end{equation}
Therefore, the magnitude of the LRTC at the interface between the domain and the domain wall ($y=0$) is equal to
\begin{equation} \label{f_Ltr}
f_L (k_n,0) \equiv f_{2L} (k_n,0) =-\frac{i f_S \cos\frac\varphi 2 \sgn\omega}{\gamma_b k_h^2} \frac Q{\kappa_Q
+\kappa_0},
\end{equation}
while the decay along the $y$ axis is determined by $\kappa_0$ or $\kappa_Q$ at $y<0$ and $y>0$, respectively [see Eq.\
(\ref{A})]. Below we consider the situations when only the LRTC is essential, and denote the corresponding contribution
$f_{2L}$ to the Green function by $f_L$ for brevity.

The real-space function is determined by the inverse Fourier transform
\begin{equation} \label{fL}
f_L (x,y) = \frac 1{2d} \sum_{k_n} e^{i k_n x} f_L(k_n,y).
\end{equation}
The $\bar f_L$ function is obtained after substituting $k_n \mapsto \bar k_n$ and $\cos\frac\varphi 2 \mapsto
\sin\frac\varphi 2$.

\subsection{Josephson current}

The supercurrent density is determined by the anomalous part of the Green function:
\begin{equation} \label{I}
\mathbf j = \frac{i \sigma\pi T}{4e} \Tr \sum_\omega \hat\tau_3 \hat\sigma_0 \check F \nabla \check F ,
\end{equation}
where $\sigma$ is the conductivity in the normal state (we shall calculate this expression inside the F layer, hence
$\sigma$ is the conductivity of the F material). Since we choose the phases of the left and right superconductor as $\mp
\varphi/2$, then the $f$ components are even with respect to the center of the interlayer, while the $\bar f$ components
are odd. Calculating the supercurrent in the center of the F layer, we obtain
\begin{equation}
\Tr \left( \hat\tau_3 \hat\sigma_0 \check F \nabla \check F \right) = 4i \left( f_0 \nabla \bar f_0 +\mathbf f \nabla
\bar{\mathbf f} \right) .
\end{equation}
Only the $x$ component of the current survives due to the odd behavior of the $\bar f$ functions.

However, calculating the current not in the center of the F layer we also obtain the $y$ component. In the region of the
F layer, where only the long-range component is present, we have $\check F = \hat\tau_1 \hat\sigma_2 f_L + \hat\tau_2
\hat\sigma_2 \bar f_L$ and finally
\begin{equation} \label{j_xy}
\mathbf j = \frac{\sigma \pi T}e \sum_\omega \left( \bar f_L \nabla f_L - f_L \nabla \bar f_L \right).
\end{equation}
In the region with the magnetization rotating as a function of $y$, the final expression remains the same with $f_L$ and
$\bar f_L$ being the second components of $\hat f_u$, see Eqs.\ (\ref{Utrans}), (\ref{f(y)}), and (\ref{delta-f}).

Inside the half-infinite domain, Eq.\ (\ref{fL}) immediately yields
\begin{align}
f_L (x,y) &= -\sum_{k_n} e^{ik_n x} \frac{i f_S \cos(\varphi/2) Q \sgn\omega}{2d\gamma_b k_h^2 (\kappa_Q +\kappa_0 )}
e^{\kappa_0 y} ,
\\
\bar f_L (x,y) &= -\sum_{\bar k_n} e^{i\bar k_n x} \frac{i f_S \sin(\varphi/2) Q \sgn\omega}{2d\gamma_b k_h^2 (\kappa_Q
+\kappa_0 )} e^{\kappa_0 y} .
\end{align}

In the limit of thin F layer, i.e., $d \ll 1/Q, 1/k_\omega$ (at the same time, we assume $1/k_h \ll d$, which allows us
to neglect the short-range components), we obtain
\begin{align}
& f_L (x,y) = -\frac{i f_S \cos(\varphi/2) \sgn\omega}{2d \gamma_b k_h^2} \notag \\
& \times \biggl( \frac Q{\sqrt{Q^2+2 k_\omega^2} +\sqrt 2 k_\omega}\; e^{\sqrt 2 k_\omega y} + \sum_{k_n>0} \frac Q{k_n}
\cos k_n x\; e^{k_n y} \biggr), \label{f_2L}
\\
& \bar f_L (x,y) = -\frac{i f_S \sin(\varphi/2) \sgn\omega}{2d \gamma_b k_h^2} \sum_{\bar k_n>0} \frac Q{\bar k_n} \cos
\bar k_n x\; e^{\bar k_n y} . \label{f_2Lbar}
\end{align}

We calculate the Josephson current according to Eq.\ (\ref{j_xy}). In the limit of thin F layer, the main contribution
is given by the second term in the brackets $f_L \nabla \bar f_L$, moreover, it is sufficient to keep only the $k_n=0$
component in $f_L$ [the first term in the brackets in Eq.\ (\ref{f_2L})]. This is because the components containing
$k_n$ in the denominator in Eq.\ (\ref{f_2L}) are much smaller. Summing over $\bar k_n$ finally yields
\begin{align}
&\mathbf j(x,y) = -\sin\varphi \frac{\sigma Q^2 \pi T}{16 e (d \gamma_b k_h^2)^2} \notag \\
&\times \sum_\omega \frac{f_S^2}{\sqrt{Q^2+2 k_\omega^2}+\sqrt 2 k_\omega} \exp\left( -\sqrt 2 k_\omega |y| \right)
\notag \\
&\times \frac{\mathbf e_x \cosh\left( \frac{\pi y}{2d} \right) \sin\left( \frac{\pi x}{2d} \right) + \mathbf e_y
\sinh\left( \frac{\pi y}{2d} \right) \cos\left( \frac{\pi x}{2d} \right)}{\sinh^2 \left( \frac{\pi y}{2d} \right) +
\sin^2 \left( \frac{\pi x}{2d} \right)} , \label{jf}
\end{align}
where $\mathbf e_x$ and $\mathbf e_y$ are the unit vectors in the $x$ and $y$ directions, respectively. In the region
with $Q\ne 0$, we obtain Eq.\ (\ref{jf}) with $\sqrt{Q^2 +2 k_\omega^2}$ instead of $\sqrt{2 k_\omega^2}$ in the
argument of the exponential. The main qualitative result of this formula is that the $x$ component of the current
density has a form $j_x = j_{cx} \sin\varphi$ with negative $j_{cx}$, i.e., the $\pi$ state of the junction is realized.

Expression (\ref{jf}) for the supercurrent is valid in the region where only the LRTC component is essential, while the
short-range components are exponentially small. Therefore, close to the SF interfaces and to the boundary between the
domain and the domain wall (at distances of the order of $\xi_h$) the expression is not applicable.

Note that $\nabla \mathbf j = 0$ and $\nabla \times \mathbf j =0$ within our accuracy (we must neglect $k_\omega$ and
$Q$ in comparison with $1/d$). The total $y$ current is absent, while the total $x$ current is equally shared between
the regions with constant and rotating magnetization and corresponds to the $\pi$ junction:
\begin{align}
&\int_0^{2d} j_y(x,y) dx = 0,
\\
&\int_{-\infty}^0 j_x(x,y) dy = \int_0^\infty j_x(x,y) dy \notag \\
&= -\sin\varphi \frac{\sigma Q^2 \pi T}{16 e d \gamma_b^2 k_h^4} \sum_\omega \frac{f_S^2}{\sqrt{Q^2+2 k_\omega^2}+\sqrt
2 k_\omega} . \label{Jc}
\end{align}

\begin{figure}
 \includegraphics[width=8cm]{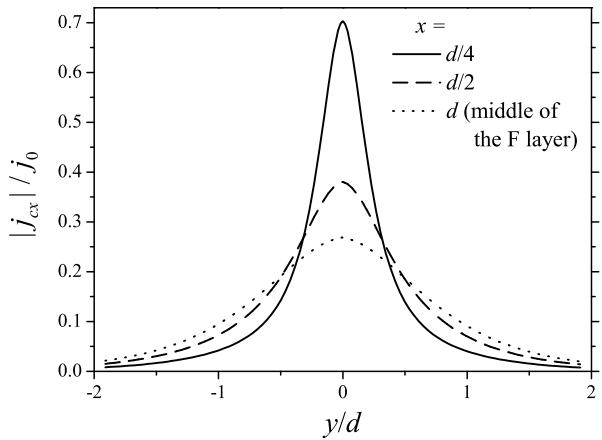}
\caption{SFS junction with a half-infinite domain: absolute value of the critical current density $j_{cx}$ in the $x$
direction as a function of the $y$ coordinate at low temperatures. The current is negative, i.e., the $\pi$ state of the
junction is realized. The normalization constant is $j_0 = \sigma D Q^3/8e (d \gamma_b k_h^2)^2$. Other parameters:
$DQ^2 = \Delta$ and $Qd=0.1$.}
 \label{fig:j_cx}
\end{figure}

The results for the $x$ component of the critical current density $j_{cx}(x,y)$ are illustrated in Fig.~\ref{fig:j_cx},
where we plot $|j_{cx}|$ as a function of $y$ at several values of $x$. The current density is maximal at the boundary
between the domain and the domain wall ($y=0$), and the maximal current across the junction is carried along this line.
At the same time, the relative height of this peak depends on $x$. The distribution is smoothest at the center of the F
layer (i.e., at $x=d$), while the peak at the domain--domain-wall boundary becomes more pronounced closer to the SF
interfaces.

In order to estimate the absolute value of the critical Josephson current following from Eq.\ (\ref{Jc}), we assume that
the junction area is $50\times 50$~$\mu$m$^2$, $\sigma \sim (50$~$\mu\Omega\;$cm$)^{-1}$, $Q\sim (50$~nm$)^{-1}$, $d\sim
\gamma_b\sim 5\xi_h$, $h\sim 500$~K, and $D\sim 10$~cm$^2$/s. Then the critical current at low temperatures is of order
3~$\mu$A which is well within the experimentally measurable range.

\section{Josephson effect in a multidomain SFS junction} \label{sec:multi}

In this section we study the LRTC in a SF structure with a multidomain ferromagnetic layer [Fig.~\ref{fig:system}(b)].
One can distinguish between two possibilities: (a)~positive chirality, when the magnetization vector $\mathbf M(y)$ in
all the domain walls rotates in the same direction (e.g., clockwise), and (b)~negative chirality, when the vector
$\mathbf M(y)$ in neighboring domain walls rotates in the opposite directions [e.g., clockwise in the $2n$th domain
walls and counterclockwise in the $(2n+1)$th domain walls]. We are interested in the LRTC assuming that the exchange
length $\xi_h$ is much smaller than the coherence length $\xi_T$. At distances $x$ essentially exceeding the length
$\xi_h$ only the LRTC survives in the F layer.

We assume that the width of the domains with $Q=0$ is $2 a_0$ and the width of the domain walls ($Q\neq 0$) is $2 a_Q$.
The origin ($y=0$) is located in the middle of a domain with the constant magnetization. At $x \gg \xi_h$ only the
long-range components of the condensate function survive in the ferromagnet. The largest long-range component is the
LRTC. At the boundary between a domain and a domain wall the solution must satisfy boundary conditions (\ref{MC1}) and
(\ref{MC2}).

\subsection{Positive chirality}

Consider first the case of positive chirality. The angle $\alpha(y)$ is then an odd function of $y$, which means that
$f_2(y)$ is also odd --- this general symmetry can be demonstrated in Eq.\ (\ref{usadelM1}). Hence the solution for the
LRTC is
\begin{align}
f_L (k_n,y) &= A \sinh (\kappa_0 y),\quad -a_0 < y<a_0, \\
f_L (k_n,y) &= B \sinh \bigl( \kappa_Q (y-a_0-a_Q) \bigr),\quad a_0< y < a_0+2a_Q .
\end{align}
Matching these solutions and their derivatives at $y= a_0$, we find
\begin{align}
B &= -A \frac{\sinh(\kappa_0 a_0)}{\sinh(\kappa_Q a_Q)} \notag \\
&=- \frac{Q \mathcal F_3}{\cosh(\kappa_Q a_Q) \left( \kappa_Q +\kappa_0 \frac{\tanh(\kappa_Q a_Q)}{\tanh(\kappa_0 a_0)}
\right)}.
\end{align}
The amplitude of the LRTC at $y=a_0$ is
\begin{equation} \label{f1a0}
f_L (k_n,a_0) =\frac{Q \mathcal F_3}{\kappa_Q \coth(\kappa_Q a_Q) +\kappa_0 \coth(\kappa_0 a_0)}.
\end{equation}

We see that $f_L (k_n,a_0)$ turns to zero both at $a_Q \to 0$ and $a_0 \to 0$. These limits mean that the widths of the
domain walls and domains are assumed to be small in comparison with $\xi_T$ but larger than $\xi_h$. The case $a_Q =0$
implies that we have a domain structure with the collinear orientation of magnetizations. The case $a_0 =0$ corresponds
to a SF structure with continuously rotating magnetization (the case studied in Ref.~\onlinecite{Eschrig2}). In both the
cases, the LRTC does not arise.

The spatial dependence of the LRTC in the domain ($|y|<a_0$), corresponding to Eq.\ (\ref{f1a0}), is given by the
inverse Fourier transformation
\begin{equation} \label{f_1Dom}
f_L (x,y) = \frac 1{2d} \sum_{n=-\infty}^\infty e^{i k_n x} f_L(k_n,a_0) \frac{\sinh (\kappa_0 y)}{\sinh(\kappa_0 a_0)}
.
\end{equation}
Interestingly, the function $f_L(x,y)$ turns to zero in the center of the domain ($y=0$). This means that the Josephson
current due to the LRTC also turns to zero in the domain center.

Formula (\ref{f_1Dom}) can be drastically simplified in the limit when the F film is thin for the long-range component
but thick for the short-range one (i.e., $k_h \gg 1/d \gg Q, k_\omega$). In this case, the main contribution is given by
the $n=0$ harmonic (with $k_n=0$), since otherwise $\kappa_Q$ and $\kappa_0$ in the denominator of Eq.\ (\ref{f1a0})
become very large. Employing also Eq.\ (\ref{F0}) we therefore obtain
\begin{equation} \label{fpc}
f_L (y) = -\frac{i f_S \cos(\varphi/2) \sgn\omega}{2d \gamma_b k_h^2} \frac{Q \sinh \left( \sqrt 2 k_\omega y
\right)}{\mathcal K_+ \sinh \left( \sqrt 2 k_\omega a_0 \right)} ,
\end{equation}
where we have denoted
\begin{gather}
\mathcal K_+ = \sqrt{Q^2 +2 k_\omega^2} \coth( \sqrt{Q^2 +2 k_\omega^2} a_Q ) \notag \\
+\sqrt 2 k_\omega \coth( \sqrt 2 k_\omega a_0 ) .
\end{gather}
The $x$ dependence has vanished since the F layer is thin and the even function $f_L(x)$ is nearly constant.

Now we can easily write down the result for $\bar f_L$, employing the rule formulated in Sec.~\ref{sec:basic}: in Eq.\
(\ref{f_1Dom}) we should substitute $k_n \mapsto \bar k_n$ and $\cos\frac\varphi 2 \mapsto \sin\frac\varphi 2$. After
that we make the final step, assuming the limit of thin F layer. This step is different from the case of the $f_L$
function, because there is no mode with $\bar k_n =0$, hence we must retain all the modes in the sum over $n$:
\begin{multline}
\bar f_L (x,y) = -\frac{i f_S \sin(\varphi/2) \sgn\omega}{2d \gamma_b k_h^2} \\
\times \sum_{\bar k_n>0} \frac{2Q \cos \bar k_n x}{\bar k_n \left[ \coth\left( \bar k_n a_Q \right) + \coth\left( \bar
k_n a_0 \right) \right]} \frac{\sinh \left( \bar k_n y \right)}{\sinh \left( \bar k_n a_0 \right)} .
\end{multline}

Finally, we find the supercurrent:
\begin{gather}
\mathbf j_0(x,y) = -\sin\varphi \frac{\sigma Q^2 \pi T}{4 e (d \gamma_b k_h^2)^2} \sum_\omega \frac{f_S^2 \sinh ( \sqrt
2 k_\omega y )}{\mathcal K_+ \sinh ( \sqrt 2 k_\omega a_0 )}
\notag \\
\times \sum_{\bar k_n >0} \frac{\mathbf e_x \sin ( \bar k_n x ) \sinh ( \bar k_n y ) - \mathbf e_y \cos ( \bar k_n x )
\cosh ( \bar k_n y )}{\left[ \coth( \bar k_n a_Q ) +\coth( \bar k_n a_0 ) \right] \sinh ( \bar k_n a_0 )} . \label{j0+}
\end{gather}
If we want to calculate the current in the region with $Q\ne 0$, then we obtain the same expression, but with $y$
counted from the center of the domain wall, with $\sqrt{Q^2 +2 k_\omega^2}$ instead of $\sqrt{2 k_\omega^2}$ in the
arguments of two $\sinh$'s, and with $a_Q$ instead of $a_0$ in the arguments of two $\sinh$'s. For example, at
$a_0<y<a_0 +2 a_Q$, we have
\begin{gather}
\mathbf j_Q(x,y) = -\sin\varphi \frac{\sigma Q^2 \pi T}{4 e (d \gamma_b k_h^2)^2} \sum_\omega \frac{f_S^2 \sinh (
\sqrt{Q^2+2 k_\omega^2} y' )}{\mathcal K_+ \sinh ( \sqrt{Q^2+2 k_\omega^2} a_Q )}
\notag \\
\times \sum_{\bar k_n >0} \frac{\mathbf e_x \sin ( \bar k_n x ) \sinh ( \bar k_n y' ) - \mathbf e_y \cos ( \bar k_n x )
\cosh ( \bar k_n y' )}{\left[ \coth( \bar k_n a_Q ) +\coth( \bar k_n a_0 ) \right] \sinh ( \bar k_n a_Q )} , \label{jQ+}
\end{gather}
where $y'=y-a_Q-a_0$.

\subsection{Negative chirality}

Consider now the case of negative chirality, when the $\mathbf M(y)$ vector rotates in the opposite directions in
neighboring domain walls. In this case the spatial dependence of $f_2 (y)$ in the domain walls remains the same as
before, i.e., this function is odd with respect to the center of a domain wall. However, the spatial dependence of the
LRTC in the domains changes drastically: it becomes an even function with respect to the center of a domain. Therefore
this dependence is
\begin{align}
f_L (k_n,y) &= C \cosh (\kappa_0 y),\quad -a_0 < y < a_0, \\
f_L (k_n,y) &= D \sinh \bigl( \kappa_Q (y-a_0-a_Q) \bigr),\quad a_0 < y< a_0+2a_Q.
\end{align}
From the boundary conditions (\ref{MC1}) and (\ref{MC2}) we find the coefficients $C$ and $D$, and finally obtain
\begin{equation} \label{f1a0_nc}
f_L (k_n,a_0) =\frac{Q \mathcal F_3}{\kappa_Q \coth(\kappa_Q a_Q) +\kappa_0 \tanh(\kappa_0 a_0)}.
\end{equation}
In this case the LRTC disappears only in the limit $a_Q \to 0$ because in this limit one again has a domain structure
with the collinear orientation of magnetizations and very narrow domain walls.

Further analysis is similar to that for the previous case of positive chirality. Inside of the domain ($|y|<a_0$) we
obtain
\begin{equation} \label{f_1Domd_nc}
f_L (x,y) = \frac 1{2d} \sum_{k_n} e^{ik_n x} f_L(k_n,a_0) \frac{\cosh (\kappa_0 y)}{\cosh(\kappa_0 a_0)} .
\end{equation}
In the limit $k_h \gg 1/d \gg Q, k_\omega$, Eq.\ (\ref{f_1Domd_nc}) yields
\begin{align}
& f_L (y) = -\frac{i f_S \cos(\varphi/2) \sgn\omega}{2d \gamma_b k_h^2} \frac{Q \cosh \left( \sqrt 2 k_\omega y
\right)}{\mathcal K_- \cosh \left( \sqrt 2 k_\omega a_0 \right)} ,
\\
& \bar f_L (x,y) = -\frac{i f_S \sin(\varphi/2) \sgn\omega}{2d \gamma_b k_h^2} \notag \\
& \times \sum_{\bar k_n>0} \frac{2Q \cos \bar k_n x}{\bar k_n \left[ \coth\left( \bar k_n a_Q \right) + \tanh\left( \bar
k_n a_0 \right) \right]} \frac{\cosh \left( \bar k_n y \right)}{\cosh \left( \bar k_n a_0 \right)} ,
\end{align}
where we have denoted
\begin{gather}
\mathcal K_- = \sqrt{Q^2 +2 k_\omega^2} \coth( \sqrt{Q^2 +2 k_\omega^2} a_Q ) \notag \\
+\sqrt 2 k_\omega \tanh( \sqrt 2 k_\omega a_0 ) .
\end{gather}
Finally, we find the supercurrent:
\begin{gather}
\mathbf j_0(x,y) = -\sin\varphi \frac{\sigma Q^2 \pi T}{4 e (d \gamma_b k_h^2)^2} \sum_\omega \frac{f_S^2 \cosh ( \sqrt
2 k_\omega y )}{\mathcal K_- \cosh ( \sqrt 2 k_\omega a_0 )}
\notag \\
\times \sum_{\bar k_n >0} \frac{\mathbf e_x \sin ( \bar k_n x ) \cosh ( \bar k_n y ) - \mathbf e_y \cos ( \bar k_n x )
\sinh ( \bar k_n y )}{\left[ \coth( \bar k_n a_Q ) +\tanh( \bar k_n a_0 ) \right] \cosh ( \bar k_n a_0 )} . \label{j0-}
\end{gather}
Similarly, in the region with $Q\ne 0$ (at $a_0<y<a_0 +2 a_Q$), we obtain
\begin{gather}
\mathbf j_Q(x,y) = -\sin\varphi \frac{\sigma Q^2 \pi T}{4 e (d \gamma_b k_h^2)^2} \sum_\omega \frac{f_S^2 \sinh (
\sqrt{Q^2+2 k_\omega^2} y' )}{\mathcal K_- \sinh ( \sqrt{Q^2+2 k_\omega^2} a_Q )}
\notag \\
\times \sum_{\bar k_n >0} \frac{\mathbf e_x \sin ( \bar k_n x ) \cosh ( \bar k_n y' ) - \mathbf e_y \cos ( \bar k_n x )
\sinh ( \bar k_n y' )}{\left[ \coth( \bar k_n a_Q ) +\tanh( \bar k_n a_0 ) \right] \sinh ( \bar k_n a_Q )} , \label{jQ-}
\end{gather}
where $y'=y-a_Q-a_0$.

\subsection{Discussion of the results}

The expressions obtained for the supercurrent are valid in the region where only the LRTC component is essential, while
the short-range components are exponentially small. Therefore, close to the SF interfaces and to the boundaries between
domains and domain walls (at distances of the order of $\xi_h$) the expressions are not applicable.

The main qualitative result of expressions (\ref{j0+}), (\ref{jQ+}), (\ref{j0-}), and (\ref{jQ-}) is that the $x$
component of the current density has a form $j_x = j_{cx} \sin\varphi$ with negative $j_{cx}$, i.e., the $\pi$ state of
the junction is realized.

The $j_x(x)$ function is even while $j_y(x)$ is odd with respect to $x=d$ (the center of the F layer) at both
chiralities. The total current in the $y$ direction is zero. Within our accuracy $\nabla \mathbf j = 0$ and $\nabla
\times \mathbf j =0$ (we must neglect $k_\omega$ and $Q$ in comparison with $\bar k_n$).

The LRTC is generated at the boundaries between the domains and the domain walls. As a result, the maximal supercurrent
in the $x$ direction is carried along these lines.

Integrating the $x$ component of the supercurrent over $y$, we find the critical current per period of the
structure:\cite{GrRyzh}
\begin{align}
J_{c+} \equiv \int_{-a_0-a_Q}^{a_0+a_Q} j_{cx+} (y) dy = -\frac{\sigma Q^2 \pi T}{4 e d \gamma_b^2 k_h^4} \sum_\omega
\frac{f_S^2}{\mathcal K_+} , \\
J_{c-} \equiv \int_{-a_0-a_Q}^{a_0+a_Q} j_{cx-} (y) dy = -\frac{\sigma Q^2 \pi T}{4 e d \gamma_b^2 k_h^4} \sum_\omega
\frac{f_S^2}{\mathcal K_-},
\end{align}
in the cases of positive and negative chirality, respectively. The total current across a junction of large area is
proportional to the number of domain walls.

\begin{figure}
 \includegraphics[width=8cm]{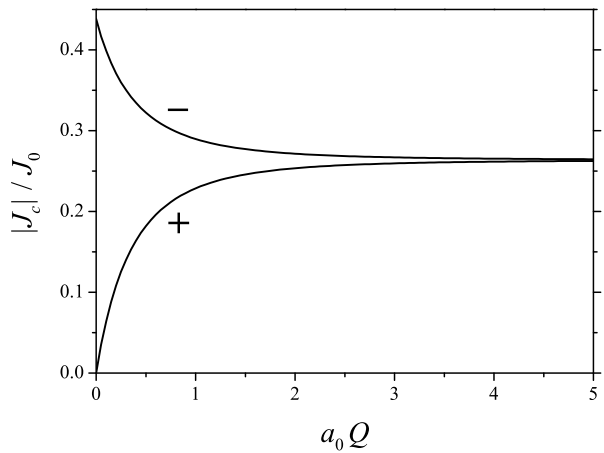}
\caption{Multidomain SFS junction: the absolute values of the critical currents $J_{c+}$ (positive chirality) and
$J_{c-}$ (negative chirality) per period of the domain structure as functions of the domain half-width $a_0$ at low
temperatures. The currents are negative, i.e., the $\pi$ state of the junction is realized. The normalization constant
is $J_0 = \sigma D Q^3/4e d \gamma_b^2 k_h^4$. Other parameters: $DQ^2 = \Delta$ and the rotation of magnetization in
the domain walls corresponds to $2 Q a_Q =\pi$.}
 \label{fig:j}
\end{figure}

The results for the critical current are illustrated in Fig.~\ref{fig:j}, where we plot $|J_{c+}|$ and $|J_{c-}|$ as
functions of the domain half-width $a_0$ (while $Q$ and $a_Q$ are linked by the condition $2 Q a_Q = \pi$ meaning
rotation of the magnetization by the angle $\pi$ in the domain wall). In the limit $a_0\to 0$ the behavior of the
supercurrent in the cases of positive and negative chiralities is drastically different. In the case of positive
chirality, the LRTC disappears in this limit and the supercurrent vanishes,\cite{comment2} while in the case of negative
chirality this is the most inhomogeneous limit and the supercurrent is maximal. In the opposite limit of large $a_0 \gg
\xi_T$, the $a_0$ dependence vanishes in both cases and the results coincide since the hyperbolic functions of $k_\omega
a_0$ turn to unity.

The appearance of the $\pi$ junction in SFS junctions is well understood in the case when it is due to the short-range
component.\cite{Buzdin} The key ingredient is the oscillating behavior of this component: the additional phase $\pi$
across the junction is provided by changing the sign. At the same time, the LRTC does not change its sign, therefore the
$\pi$ junction due to the LRTC seems counterintuitive. Where does the additional phase come from? Note that the LRTC in
our case is purely imaginary [see, e.g., Eq.\ (\ref{fpc})]. This means that there is a $\pi/2$ phase rotation at the SF
interfaces, and the two interfaces provide the $\pi$ shift. The mechanism of the $\pi$ junction due to $\pi/2$ interface
shifts is similar to Ref.~\onlinecite{GKF}.

Another type of SF structures, sensitive to the chirality of the vector $\mathbf M$, was considered in
Refs.~\onlinecite{VBE} and~\onlinecite{BVEmanif}. It was shown that the sign of the critical Josephson current in a
multilayered SF structure depends on chirality. Similarly to the present paper, the $\pi$ junction was found in the
situation when only the LRTC is essential.

\section{DOS in SF bilayer} \label{sec:dos}

\begin{figure}
\includegraphics[width=7cm]{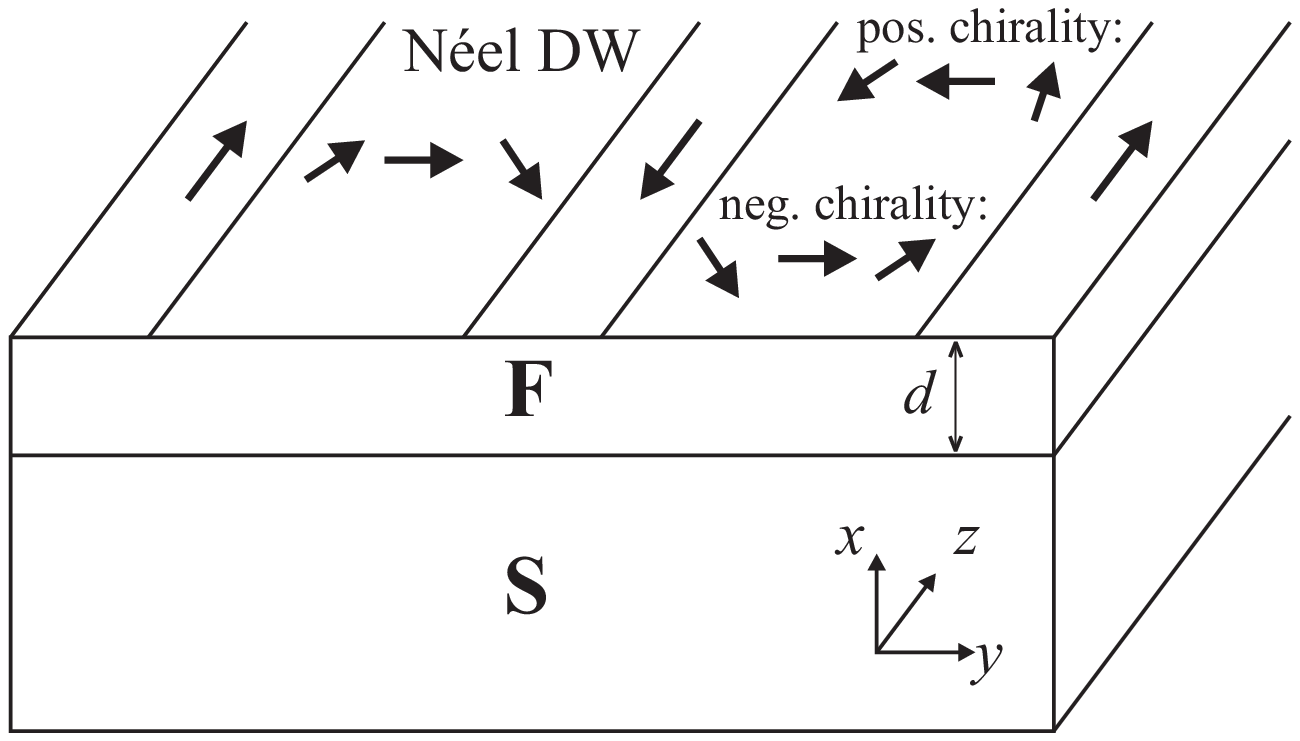} \caption{Multidomain F layer of thickness $d$ in contact with a bulk
superconductor. Depending on the relative orientation of rotating magnetizations in neighboring domain walls, we
distinguish the cases of positive and negative chirality ($Q$ has the same or opposite sign in the neighboring domain
walls, respectively). The proportion between the widths of domains and domain walls is chosen only for drawing
purposes.}
 \label{fig:system_dos}
\end{figure}

Our previous paper, Ref.~\onlinecite{VFE}, was devoted to studying the density of states (DOS) at the free surface of
the F layer in the system shown in Fig.~\ref{fig:system_dos} (with F layer of thickness $d$). These results are
immediately reproduced from the Green function calculated in the center of the F layer in the SFS junction of
Fig.~\ref{fig:system}(b) (with F layer of thickness $2d$) at zero phase difference, $\varphi=0$. Making analytical
continuation from the Matsubara frequency $\omega$ to the real energy $\varepsilon$, we obtain the correction to the DOS
due to the proximity effect as
\begin{equation} \label{DOSa}
\delta \nu (\varepsilon) = - \left. \frac{\Real f_L^2}2 \right|_{\omega \rightarrow -i\varepsilon}
\end{equation}
(we consider the region in space where only the LRTC is essential).

\begin{figure}
 \includegraphics[width=8cm]{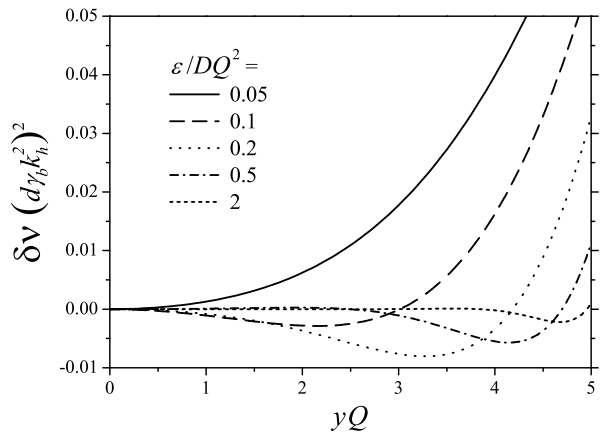}
\caption{Addition to Fig.~2 from Ref.~\onlinecite{VFE}: Correction $\delta\nu(y)$ (due to the proximity effect) to the
DOS at the free surface of the F layer in the case of positive chirality. The curves are plotted at several energies
$\varepsilon$. The width of the domains is $a_0 = 5/Q$, while the rotation of magnetization in the domain walls
corresponds to $Q a_Q =\pi$.}
 \label{fig:DoS_y}
\end{figure}

\begin{figure}
 \includegraphics[width=8cm]{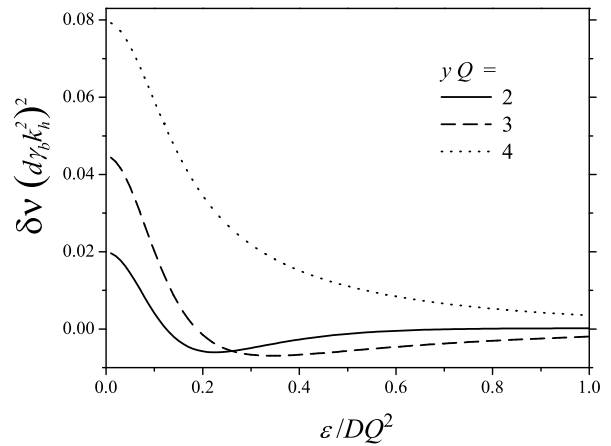}
\caption{Correction $\delta\nu(\varepsilon)$ to the DOS at the free surface of the F layer in the case of positive
chirality. The curves are plotted at several points $y$. Other parameters are the same as in Fig.~\ref{fig:DoS_y}.}
 \label{fig:DoS_e}
\end{figure}

We want to return to this question in view of the recent paper Ref.~\onlinecite{TG}, where it was demonstrated that
general analytical properties of the Green function imply that if the superconductivity has odd frequency symmetry, then
$\delta \nu(0) >0$. The expressions for the DOS from our paper Ref.~\onlinecite{VFE} testify that this general statement
is satisfied in our case, however, this fact was not illustrated in the figures. Figures~\ref{fig:DoS_y}
and~\ref{fig:DoS_e} supplement the figures from Ref.~\onlinecite{VFE} and demonstrate that $\delta \nu(0) >0$.

Figure~\ref{fig:DoS_y} is plotted for the same parameters as Fig.~2 in Ref.~\onlinecite{VFE} and shows the spatial
dependence of the DOS inside a domain ($y$ is counted from the center of the domain) at several energies. Indeed, the
DOS at low energies becomes positive everywhere.

Figure~\ref{fig:DoS_e} illustrates the $\delta\nu(\varepsilon)$ dependence at several points $y$. In accordance with
Ref.~\onlinecite{TG}, the zero-energy correction to the DOS is positive forming the zero-energy peak.

\section{Discussion} \label{sec:discuss}

In our study we neglected orbital effects of the magnetic field in the ferromagnet. Below we demonstrate that this is
justified for thin F films. We also estimate the influence of the spin-orbit interaction on the LRTC. Finally, we
comment upon the role of the boundary conditions at the SF interfaces in our problem.

The magnetization $\mathbf M$ in the F layer leads to the appearance of the vector potential $\mathbf A$ which can be
estimated as $A \sim 4\pi M d$. The vector potential enters Eq.\ (\ref{usadelM}) as an additional term $(2 \pi
A/\phi_0)^2$, where $\phi_0 =hc/2e$ is the magnetic flux quantum. This term will restrict the penetration length of the
LRTC if it is larger than the term $2 k_\omega^2 = 2 |\omega|/D \sim 2 \pi T/D$. In the opposite limit
\begin{equation} \label{Orbital}
\left( \frac{8\pi^2 M d}{\phi_0} \right)^2 < \frac{2 \pi T}D
\end{equation}
one can neglect the orbital effects (the effect of Meissner currents on the LRTC). Taking $M \sim 50$~G, $T \sim 5$~K,
and $D \sim 10$~cm$^2$/s, we obtain $d< 300$~nm. Therefore the orbital effects can be neglected in the case of thin F
layers. One can also show that under these conditions the Meissner currents induced in the superconductors by the stray
magnetic fields are much smaller than the depairing currents. Therefore one can neglect the influence of the magnetic
field of the ferromagnet on the amplitude of the order parameter in the superconductors. There is one more effect of the
domain structure in the ferromagnet. This effect occurs in Josephson junctions with lateral dimensions larger than the
Josephson length $\lambda_J$ and is related to a spatial variation of the phase difference. Due to this a modulation of
the total critical Josephson current arises. This effect was studied theoretically in Ref.~\onlinecite{Volkov06}.

If the spin-orbit interaction is present in the F layer, it leads to a decrease of the LRTC penetration
length.\cite{BVE_review} In this case the wave vector $k_\omega^2$ should be replaced by $k_\omega^2 + 4/
\tau_\mathrm{so} D$, where $\tau_\mathrm{so}$ is the spin-orbit relaxation time. Thus, the LRTC penetration length
cannot be larger than $\sqrt{\tau_\mathrm{so} D /8}$.

Note also that we use boundary conditions (\ref{BC}) which follow from the Kupriyanov-Lukichev ones and describe
spin-conserving interfaces with potential barriers. These boundary conditions themselves do not lead to the appearance
of the LRTC in the uniform ferromagnet. In some papers spin-active interfaces were studied.\cite{Millis,Zhao} The
scattering at such interfaces can generate the LRTC even in the uniform ferromagnet (see Ref.~\onlinecite{Eschrig06} and
references therein). The boundary conditions for the spin-active interfaces were written in terms of a phenomenological
scattering matrix. The microscopic calculation of the scattering matrix is lacking and the properties of spin-active SF
interfaces in experiment are not known at present.

In a recent paper Ref.~\onlinecite{Asano} the LRTC was analyzed numerically in a model of a SFS structure with a
half-metallic ferromagnet and spin-active interfaces. The exchange field near the SF interfaces was assumed to be
inclined with respect to the exchange field in the bulk of the ferromagnetic layer. To some extent this model is similar
to a model of SF structure with a domain wall at the SF interface analyzed in Ref.~\onlinecite{BVElong}. Therefore, the
spin-active interfaces are additional sources of the LRTC in the ferromagnet. In order to single out the effect of the
N\'eel domain walls on the LRTC, we did not consider the spin-active interfaces.

\section{Conclusions} \label{sec:concl}

We have studied a Josephson junction between two superconductors through a multidomain ferromagnet (F) with an in-plane
magnetization, assuming that the neighboring domains are separated by the N\'eel domain walls. Due to an inhomogeneous
magnetization, the long-range triplet superconducting component (LRTC) arises in the system. Arising at the domain
walls, this component spreads into domains over a long distance of the order $\xi_T = \sqrt{D/2\pi T}$, which is much
greater than the usual short length $\xi_h = \sqrt{D / h}$ describing the decaying of superconducting correlations in a
ferromagnet with a homogeneous magnetization.

We have calculated the Josephson current due to this component in the case when the short-range components exponentially
decay over the thickness of the F layer and can be neglected. Focusing on the limit when the F layer is thin from the
viewpoint of the long-range component we find that the junction is in the $\pi$ state. The LRTC does not oscillate
inside the F layer, while the additional $\pi$ phase of the condensate wave function is due to $\pi/2$ shifts at the SF
interfaces. This interpretation suggests that the junction must be in the $\pi$ state due to the LRTC regardless of the
F layer's thickness. When the F layer is not thin (from the viewpoint of the LRTC), analytical expressions for the
supercurrent become cumbersome, however, numerical calculations indicate that the junction is indeed in the $\pi$ state.

The absolute value of the Josephson current density is maximal at the boundaries between domains and domain walls, see
Fig.~\ref{fig:j_cx}. The current mainly flows along these lines.

We have considered two possible chiralities of the domain structure (positive and negative), which are determined by the
relative orientation of rotating magnetizations in the neighboring domain walls. The absolute value of the Josephson
current due to the LRTC is larger in the case of the negative chirality, because this case corresponds to a more
inhomogeneous magnetization.

Analyzing a correction to the density of states due to the LRTC, we find that at zero energy (i.e., at the Fermi level)
the correction is positive. This fact is in accordance with the general statement from Ref.~\onlinecite{TG}, based on
the odd frequency dependence of the Green function.

\begin{acknowledgments}
We are grateful to Y.\ Tanaka for helpful discussions. We would like to thank SFB 491 for financial support. Ya.V.F.\
was also supported by RFBR Grant No.\ 04-02-16348, RF Presidential Grants Nos.~MK-3811.2005.2 and MK-4421.2007.2, the
Dynasty Foundation, the program ``Quantum Macrophysics'' of the RAS, CRDF, and the Russian Ministry of Education.
A.F.V.\ also thanks DFG for financial support within Mercator-Gastprofessoren.
\end{acknowledgments}


\begin{thebibliography}{99}
\bibitem{comment2}
Actually, the supercurrent in the limit $a_0\to 0$ does not vanish exactly, but is determined by an exponentially small
contribution from the short-range component which we neglect in this paper.
\end{thebibliography}
\end{document}